\documentclass{ws-procs975x65}

\pdfoutput=1                                  

\usepackage{lineno}
\usepackage{graphicx}
\usepackage{amsmath}   
\usepackage{subfigure}

\begin{document}
\title{CALET: a high energy astroparticle physics experiment on the ISS}
\author{Pier S. Marrocchesi$^*$ (for the CALET collaboration)}
\address{Dept.$\,$of Physical Sciences, Earth and Environment, Via Roma 56, I-53100 Siena, Italy\\
and INFN Sezione di Pisa, Largo Bruno Pontecorvo 3, I-56127 Pisa, Italy\\
$^*$E-mail: piersimone.marrocchesi@pi.infn.it\\
}

\vspace{-12pt}
\begin{abstract}
CALET (CALorimetric Electron Telescope) is a high energy astroparticle physics experiment planned for a long exposure mission aboard the International Space Station (ISS) by the Japanese Aerospace Exploration Agency, in collaboration with the Italian Space Agency (ASI) and NASA. 
The main science goal is high precision measurements of the inclusive electron ($\scriptstyle+$positron) spectrum below 1 TeV and the exploration of the energy region above 1 TeV, where the shape of the high end of the spectrum might unveil the presence of nearby sources of acceleration.  CALET has been designed to achieve a large proton rejection capability ($>$10$^5$) with a fine grained imaging calorimeter (IMC) followed by a total absorption calorimeter (TASC), for a total thickness of 30 X$_{0}$ and 1.3 proton interaction length. With an excellent energy resolution and a lower background contamination with respect to previous experiments, CALET will search for possible spectral signatures of dark matter with both electrons and gamma rays.
CALET will also measure the high energy spectra and relative abundance of cosmic nuclei from proton to iron and will detect trans-iron elements up to  Z$\,\sim$\,40.  The charge identification of individual nuclear species is performed by a dedicated module (CHD), at the top of the apparatus, and by multiple dE/dx measurements in the IMC. 
With a large exposure and high energy resolution, CALET will be able to verify and complement the observations of CREAM, PAMELA and AMS-02 on a possible deviation from a pure power-law of proton and He spectra in the region of a few hundred GeV and to extend the study to the multi-TeV region. CALET will also contribute to clarify the present experimental picture on the energy dependence of the boron/carbon ratio, below and above 1 TeV/n, thereby providing valuable information on cosmic-ray propagation in the galaxy.
Gamma-ray transients will be studied with a dedicated Gamma-ray Burst Monitor (GBM). 
\end{abstract}

\vspace{-5pt}
\bodymatter

\vspace{-1pt}
\section{Introduction}

The CALorimetric Electron Telescope (CALET) is a space based experiment for long-term observations of high-energy cosmic radiation on the International Space Station (ISS).  CALET is an all-calorimetric instrument \cite{Torii2011, Torii2013} designed to achieve a large proton rejection capability ($>$10$^{5}$) with a fine grained imaging calorimeter (IMC) followed by a total absorption calorimeter (TASC), for a total thickness of 30 X$_{0}$ and $\sim$1.3 proton interaction length ($\lambda_{I}$). 
The charge identification of individual nuclear species is performed by a two-layered hodoscope of plastic scintillators (CHD) at the top of the apparatus  (Fig.1a), providing a measurement of the charge Z of the incident particle over a wide dynamic range (Z$\,$=$\,$1 to $40$) with sufficient charge resolution to resolve individual elements \cite{Shimizu2011, Mar2011} and complemented by a redundant charge determination via multiple dE/dx measurements in the IMC. 
The IMC is a sampling calorimeter, composed of 16 layers of scintillating fibers with 1 mm$^2$ squared cross-section
arranged along orthogonal directions and interspaced with thin tungsten absorbers.
It can image the early shower profile in the first 3 X$_{0}$ and reconstruct the incident direction of cosmic rays with good angular resolution.
The TASC is a 27 X$_0$ thick homogeneous calorimeter with 12 alternate X-Y layers of lead-tungstate (PWO) logs. It measures the total energy of the incident particle and discriminates electrons from hadrons with the help of the information from the CHD and IMC.
The instrument is described in more detail elsewhere \cite{Torii2011, Torii2013, Mar2012}.\\
The CALET mission carried out preliminary phase studies in 2007-09, followed by the construction and commissioning of the payload, leading to a successful launch of the instrument on 19 August 2015 from the Tanegashima Space Center (Japan). 
CALET reached the ISS on 24 August on board of the transfer vehicle (Kounotori-5) and was emplaced on the Exposure Facility of the Japanese Experimental Module (JEM-EF).
At the beginning of October the preliminary phase of on-orbit check-out and calibrations were accomplished. 
At the time of writing, the instrument is operating in science data mode transmitting data to the ground stations.\\

\vspace{-26pt}
\begin{figure}[!h]
\centering
\includegraphics[width=0.43\columnwidth]{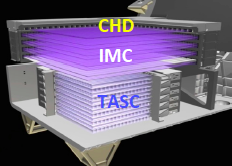}
\includegraphics[width=0.53\columnwidth]{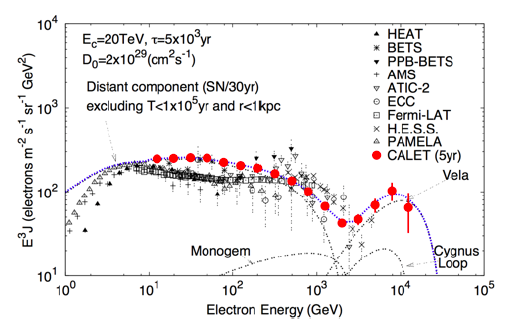}
\caption{\scriptsize
(a) left: Cross-section of the CALET instrument. From top to bottom: CHD hodoscope, IMC imaging calorimeter and TASC total-absorption calorimeter;
(b) right: Expected electron spectrum data points (in red) from CALET above 1 TeV in 5 years of observations in the hypothetical case of a prominent spectral contribution from Vela.}
\label{fig:raya}
\end{figure}

\vspace{-25pt}
\section{Main science goals}
\vspace{-5pt}
The CALET telescope will perform precise measurements of high energy cosmic rays over a target period of five years,
with an extensive physics program that includes the detection of possible nearby sources of high energy electrons; searches for signatures of dark matter in the spectra of electrons and $\gamma$-rays; long exposure observations of cosmic nuclei from proton to iron and trans-iron elements;
measurements of the CR relative abundances and secondary-to-primary ratios; monitoring of gamma-ray transients 
and studies of solar modulation.\\

\vspace{-24pt}
\section{The electron spectrum}
\vspace{-3pt}
The primary science goal of CALET \cite{Torii2015} is to perform high precision measurements of the electron spectrum from 1 GeV to 20 TeV. CALET will  scan very accurately the energy region already covered by previous experiments, taking advantage of an excellent energy resolution and a low background contamination. By integrating a sufficient exposure on the ISS, CALET will be able to explore the energy region above 1 TeV where the presence of nearby sources of acceleration is expected to shape the high end of the electron spectrum and leave faint, but detectable, footprints in the anisotropy. In order to meet this experimental goal, CALET has been designed to achieve a large proton rejection capability ($>$10$^{5}$) thanks to a full containment of electromagnetic showers in the calorimeter and a fine-grained imaging of their early development in a 3 X$_{0}$ thick pre-shower (IMC).  
\\
\underline{The TeV region}.
An exciting possibility is that the observation of the electron spectrum in the TeV region may result in a direct detection of nearby astrophysical sources of high energy electrons. In fact, the most energetic galactic cosmic-ray (GCR) electrons that can be observed from Earth are likely to originate from sources younger than $\sim$10$^5$ years and located at a distance less than 1 kpc from the Solar System. This is due to the radiative energy losses that limit the propagation lifetime of high energy electrons and, consequently, the distance they can diffuse away from their source(s).
Since the number of potential sources satisfying the above constraints are very limited, the energy spectrum of electrons might have a distinctive structure \cite{Kobayashi2004}
and the arrival directions are expected to show a detectable anisotropy.
There are at least nine candidate Supernova Remnants (SNR) with ages $<$ 10$^5$ years and distances less than 1 kpc from the solar system. Possible contributions to the observed GCR electron spectrum from both distant and nearby sources were calculated. Known candidates that may give a contribution in the TeV region include Vela, Cygnus loop and Monogem, in order of strength. Among these, Vela is very promising (Fig.1b) as both the distance ($\sim$ 0.25 kpc) and the age ($\sim$ 10$^4$ years), are very suitable for the observation.
The TeV region might as well conceal a completely different scenario
where "nearby" acceleration sources would not be detected and the spectrum found to roll off at a characteristic cutoff energy.  In this case, the measurement of the "end point" of the electron spectrum could be used to constrain the cosmic-ray diffusion coefficient.
\\
\underline{The sub-TeV region.}
The electron energy spectrum from 10 GeV to 1 TeV could be the result of the contribution of several unresolved sources.
In this 
 region CALET accuracy and long exposure will allow to significantly improve the knowledge of the detailed spectral shape and angular distribution of the inclusive electron $\scriptstyle+$ positron spectrum. This will provide information on the average features of the source spectrum, the diffusion time, the density of sources and possibly their nature, either as astrophysical objects (e.g. a nearby pulsar) or the result of the annihilation/decay of dark matter particles \cite{Motz2015}. Both possibilities have been proposed to interpret the recent measurements suggesting a hardening of the inclusive spectrum in the range 200 GeV - 1 TeV.  The presence of an additional spectral component is also required to explain the now established rise of the positron fraction above $\sim$10 GeV as measured by PAMELA \cite{Pamela-elettroni} and extended to the hundreds GeV region by AMS-02\cite{AMS02-positroni}.\\

\vspace{-16pt}
\section{Cosmic-ray spectra}  
\vspace{-5pt}

CALET will also perform long term observations of light and heavy cosmic nuclei from proton to iron and will also detect trans-iron elements up to
 Z$\,\sim\,$40. It will be able to identify the most abundant CR elements with individual element resolution and measure their spectral shape and relative abundance in the energy range from a few tens of GeV to several hundreds of TeV \cite{Mar2012}. 
CALET will first investigate - with very high accuracy - the intermediate energy region from 200 GeV/n to 800 GeV/n where a deviation from a single power-law has been reported for both proton and helium spectra by CREAM \cite{CREAMDiscrepant2010} and PAMELA \cite{Adriani2011} and recently confirmed with high statistics measurements by AMS-02\cite{AMS15_prot}.  CALET will carry out an accurate scan of this energy
region to verify the presence of a progressive hardening of the spectrum by measuring its curvature and spectral break point position. In a relatively short time, CALET will be able to close the gap between the AMS-02 highest energy points and CREAM lowest points for proton and He, extending the reach of precision measurements to the multi-TeV region. 
An example is given in Fig.2a where the expected proton data points (filled red circles) from CALET after 1 year of data taking are calculated, assuming the AMS-02 spectral parametrization and taking into account the expected efficiencies.
The errors are statistical only and refer to a restricted fiducial acceptance corresponding to a geometric factor of $\sim$0.04 m$^{2}$sr (about 1/3 of the whole acceptance). 
With AMS-02 momentum measurements limited\footnote{AMS-02 energy range can be extended for nuclei with Z $>$ 2 by using the TRD, while for protons and He the calorimeter's energy measurement is limited by an equivalent thickness of $\sim$0.5 interaction length that significantly reduces the expected number of proton interactions. } to a few TV by its MDR, precise observations of the proton and helium fluxes in the multi-TeV region are likely to come from purely calorimetric experiments already in orbit like CALET, or missions scheduled for a launch in the near future, like DAMPE \cite{Chang2013}
and ISS-CREAM \cite{ISSCREAM2015}.
On a longer observation time scale of 5 years, CALET is expected to explore the proton energy spectrum up to $\sim$900 TeV, the He spectrum to $\sim$400 TeV/amu and to measure the energy spectra of the most abundant heavy nuclei with sufficient statistical precision up to $\sim$20 TeV/amu for C and O and $\sim$10 TeV/amu for Ne, Mg, Si and Fe.  \\



\vspace{-24pt}
\begin{figure}[!h]
\centering
\includegraphics[width=0.49\columnwidth]{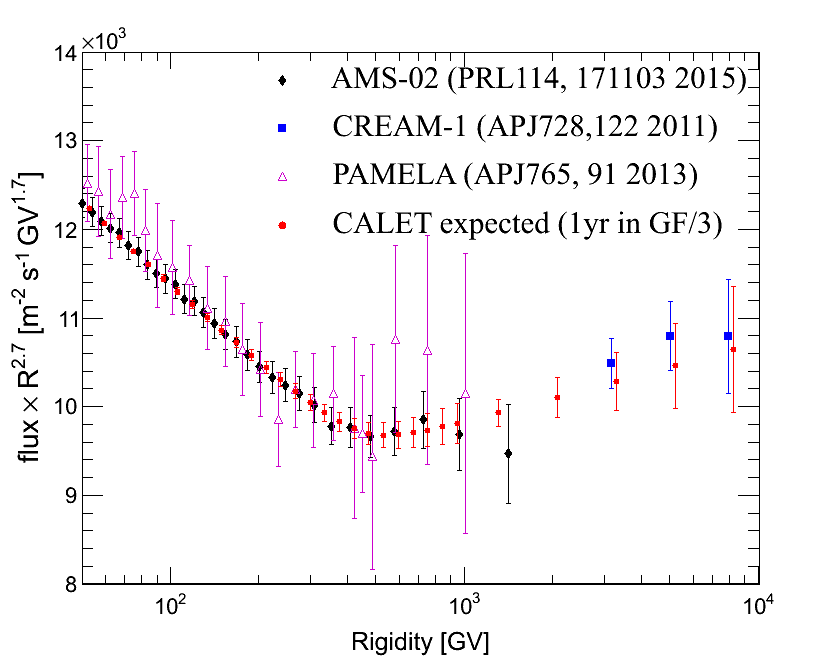}
\includegraphics[width=0.49\columnwidth]{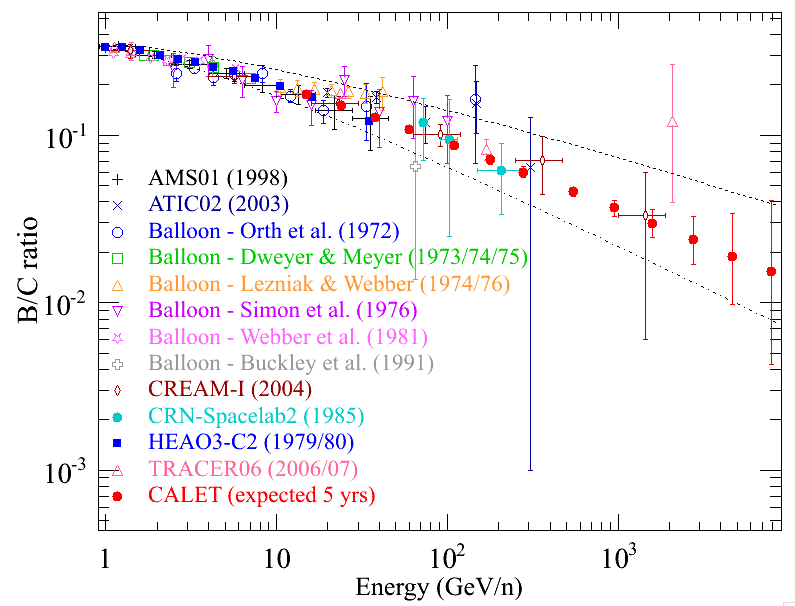}
\caption{\scriptsize (a) left: Proton rigidity spectrum from 50 GeV to 10 TV: AMS-02 \cite{AMS15_prot} data points (diamonds); PAMELA  (open triangles) 
data points (from \cite{Adriani2011} lowered by 3.2\% as prescribed in \cite{PAMELA-protoni-2013}); CALET (red filled circles) as expected after 1 year of observations in a restricted ($\sim1/3$) fiducial acceptance (statistical errors only) and CREAM-I data points below 10 TeV \cite{CREAM-1}; (b) right: A partial compilation of B/C data including CALET expected data (red points) after 5 years. The dashed (dot-dashed) lines are drawn for a Leaky Box Model with -$\delta$ = 0.33, 0.50, respectively.}
\label{fig:raya}
\end{figure}

\vspace{-12pt}
\underline{Secondary-to-primary flux ratios.}
Direct measurements of the energy dependence of the flux ratio of secondary-to-primary elements (e.g.: Boron/Carbon, sub-Fe/Fe) can discriminate among different models of CR propagation in the galaxy.  This observable is less prone to systematic errors than absolute flux measurements. Above 10 GeV/amu, the energy dependence of the propagation pathlength is often parametrized in the form E$^{- \delta}$. An accurate measurement of the spectral index parameter $\delta$ is crucial to derive the spectrum at the source by correcting the observed spectral shape for the energy dependence of the propagation term. These measurements have been pushed to the highest energies with Long Duration Balloon (LDB) experiments, however, at present, they remain statistics limited to a few hundred GeV/amu and suffer from a systematic uncertainty, due to the production of secondary nuclei in the residual atmospheric grammage at balloon altitude, that may become dominant in the TeV/amu region.
With a long exposure and in the absence of atmosphere, CALET can provide new data to improve the accuracy of the present measurements of the B/C ratio above 100 GeV/amu and extend them beyond 1 TeV/amu. A compilation of B/C data from direct measurements is shown in Fig.2b,
 where the data points expected from CALET in 5 years are marked as red filled circles in the energy range from 15 GeV/n to $\sim$8 TeV/n.\\
 
\vspace{-25pt}
\section{Dark Matter searches and gamma-ray astrophysics.}
\vspace{-5pt}
Dark Matter (DM) candidates
include WIMPs (Weakly Interacting Particles) from supersymmetric theories, such as the LSP neutralino, that may annihilate and produce gamma rays and positrons as a signature. CALET will perform a sensitive search of DM candidates in the inclusive electron spectrum, as discussed above, and in gamma-ray spectra. 
According to a class of models, the annihilation/decay of dark-matter particles in the galactic halo could produce sharp gamma-ray lines in the sub-TeV to TeV energy region, superimposed to a diffuse photon background. CALET will be capable of investigating such a distinctive signature, thanks to a gamma-ray energy resolution of 3\% above 100 GeV, that can be improved to 1\% with a reduced on-axis effective area (fiducial volume acceptance cuts to require a total lateral containment of the shower).
The precise determination of the line shape of any spectral feature is expected to play a crucial role to discriminate among different models of dark matter, or it might suggest an alternative astrophysical interpretation.
Another class of DM candidates, as suggested in \cite{Feng2002}, are Kaluza-Klein (KK) particles resulting from theories involving compactified extra-dimensions. They may annihilate in the galactic halo and produce an excess of positrons observable at Earth. Unlike neutralinos, however, direct annihilation of KK particles to leptons is not suppressed and, consequently, the KK electron “signal” is enhanced relative to that from neutralinos.
A sharp cutoff in the excess positrons close to the KK mass, might produce a detectable “feature” in the inclusive electron/positron energy spectrum.
Dark matter KK particles can also decay into gamma rays and a difference in the line shape between a neutralino and a KK candidate has to be expected \cite{Bergstrom2006}.  With a better energy resolution than Fermi-LAT above 10 GeV, CALET could have the best capability to resolve the nature of dark matter by studying the shape of any possible high energy gamma ray "line" that might be observed.
\\
\underline{Gamma-ray sources.}
Observation of gamma-ray sources will not be a primary objective for CALET.  However, its excellent energy resolution and good angular resolution (better than 0.4$^{\circ}$, including pointing uncertainty) will allow for accurate measurements of diffuse gamma-ray emission and detection of more than 100 bright sources at high latitude from the Fermi-LAT catalogue. Given the on-axis effective area of $\sim\,$600 cm$^2$ for energies above 10 GeV (reduced by $\sim\,$50\% at 4 GeV) and field of view of 45$^{\circ}$ from the vertical direction, CALET is expected to detect $\sim\,$2.5$\times 10^{4}$ ($\sim\,$7000) photons from the galactic (extra-galactic) background with E $>$ 4 GeV and $\sim\,$300 photons from the Vela pulsar with E $>$ 5 GeV.
\\
\underline{Gamma-ray Transients.}
CALET will also monitor X-ray/ gamma-ray transients in the energy region 7 keV to 20 MeV with a dedicated Gamma-ray Burst Monitor (CGBM). It will extend GRB studies carried out by other experiments (e.g. Swift and Fermi/LAT) and provide added exposure when the other instruments will not be available or pointing to a different direction.  Furthermore, high energy photons possibly associated with a burst event can be recorded over the entire CALET energy range down to 1 GeV where the CALET main telescope has still (limited) sensitivity, albeit with low resolution. Upon the detection of a GRB, an alert will be transmitted to a network of ground "antennas" (including LIGO and VIRGO) for the possible simultaneous detection of gravitational waves associated with the event.

\vspace{-15pt}
\section{Conclusions}
\vspace{-2pt}
CALET reached the ISS on August 24, 2015
and was emplaced on the Exposure Facility JEM-EF.
%
At the time of writing, the instrument is operating in science data mode transmitting data to the ground stations.
A 2 years period of observations has started with a target of 5 years.\\



\end{document}